\documentclass[article]{revtex4}

\usepackage{graphics}
\usepackage{graphicx}
\usepackage{bm}
\usepackage{latexsym}
\usepackage{color}
\usepackage{mathrsfs}
\usepackage{braket}
\usepackage{enumerate}
\usepackage[usenames,dvipsnames]{xcolor}
\usepackage[colorlinks=true,citecolor=blue,linkcolor=blue,urlcolor=blue,backref=true]{hyperref}
\usepackage{float}
\usepackage[raggedright,tight]{subfigure}  %side-by-side figures
\usepackage{amsmath}
\usepackage{amsfonts}

\begin{document}

\title{Quantum optical dipole radiation fields}
\author{Adam Stokes}
%\address{The School of Physics and Astronomy, University of Leeds, Leeds LS2 9JT, United Kingdom} 
%\ead{a.stokes@leeds.ac.uk}

\begin{abstract}
We introduce quantum optical dipole radiation fields defined in terms of photon creation and annihilation operators. These fields are identified through their spatial dependence, as the components of the total fields that survive infinitely far from the dipole source. We use these radiation fields to perturbatively evaluate the electromagnetic radiated energy-flux of the excited dipole. Our results indicate that the standard interpretation of a bare atom surrounded by a localised virtual photon cloud, is difficult to sustain, because the radiated energy-flux surviving infinitely far from the source contains virtual contributions. It follows that there is a clear distinction to be made between a radiative photon defined in terms of the radiation fields, and a real photon, whose identification depends on whether or not a given process conserves the free energy. This free energy is represented by the difference between the total dipole-field Hamiltonian and its interaction component.
\end{abstract}

\maketitle

\section{Introduction}

As the first successful field theory Maxwell electrodynamics revolutionised our understanding of light and matter, and its wide applicability has ensured its status as one of the pillars of modern physics. At the same time its beauty and elegance has resulted in a plethora of theoretical advancements and generalisations. Its basic ontological component is the electromagnetic field, which assigns electromagnetic properties to events in spacetime. The source of these fields is charged matter, and the behaviour of the fields can be classified in terms of the distance from the source. In classical Maxwell theory radiation is typically defined as electromagnetic energy that survives infinitely far away from the source, and is therefore able to ``detach" itself from the source in question. The radiation fields are defined as the components of the total fields that contribute to this radiated energy. The remainder of the electromagnetic energy is viewed as permanently attached to the charged source (see, for example, \cite{griffiths_introduction_1999}).

In Quantum electrodynamics and quantum optics the primary ontological object is the photon. Photons are discrete quanta of electromagnetic energy, but aside from their discrete nature, they do not behave like particles in any conventional sense. As relativistic quanta, they cannot be localised in a way analogous to the non-relativistic wave-mechanical particle.

The primary theoretical predictions in quantum electrodynamics come in the form of $S$-matrix elements and cross-sections calculated using time-dependent perturbation theory. In such calculations the Hamiltonian $H$ is split into free and interaction components $H=H_0+V$. For a single dipole within the Maxwell field the free component can be further partitioned as $H_0=H_D+H_F$ where $H_D$ and $H_F$ depend solely on the canonical operators of the dipole and field respectively. The reason for the partitioning $H=H_0+V$ is that the free component $H_0$ can often be diagonalised exactly, its eigenstates being referred to as bare states. \emph{Processes} in QED are then understood as transitions between bare states that occur when the interaction $V$ is switched on adiabatically from the remote past $t=-\infty$ and switched off adiabatically in the distant future $t=+\infty$.

Photons are often classified as real or virtual based on the extent to which the associated emission event conserves the free energy $H_0$, and phenomena are then interpreted in terms of either real or virtual photons. Spontaneous emission, for example, refers to the irreversible emission of real photons. Level-shift phenomena, on the other hand, are interpreted as resulting from interactions involving virtual photons.

The conceptual differences between classical and quantum theory with regards to the understanding of radiative phenomena are of continual interest, and the transition between the two theories has been considered extensively \cite{heitler_quantum_1954,power_introductory_1965,cohen-tannoudji_photons_1989,milonni_quantum_1994,novotny_principles_2006}. In \cite{erber_unified_2003} the authors use covariant Fourier transforms to provide a framework which to some extent unifies classical and quantum theories for the treatment of the angular distribution of radiation. In \cite{andrews_virtual_2004} the role of virtual contributions as part of a unified quantum electrodynamic treatment of energy transfer between space-like separated dipoles is discussed. This analysis offers a clarification of the distinction between radiative and radiationless contributions to energy transfer. The nature of energy transfer, causality and the virtual field, have also received wide-spread attention in the context of the famous Fermi two-atom problem \cite{biswas_virtual_1990,hegerfeldt_causality_1994,buchholz_there_1994,milonni_photodetection_1995,power_analysis_1997,sabin_fermi_2011,stokes_noncovariant_2012}. In \cite{rice_identifying_2012} the authors analyse the spatial dependence of transition matrix elements of the quantum dipolar interaction Hamiltonian in such a way that exposes the varying character, other than simply varying light intensity, of electromagnetic phenomena in the near, intermediate and far zones.

Here, we investigate virtual photon contributions made by the quantum versions of the classical radiation fields. These virtual contributions arise in the interacting setting, because the free vacuum $\ket{0}$, which is the ground state of $H_0$, does not coincide with the interacting ground state of $H$ \cite{kurcz_energy_2010,kurcz_rotating_2010,stokes_extending_2012}. The interaction can therefore result in the production of photons from the vacuum state $\ket{0}$ with the simultaneous excitation of the dipole. Similarly, the dipole can absorb photons from an excited state and make a transition into a lower energy state. Clearly such processes do not conserve the free energy, and moreover they tend to occur over very short timescales. They are distinct from the vacuum-fluctuations associated with the non-vanishing of the vacuum expectation values of the squares of the Maxwell fields, the latter already occurring within free quantum electrodynamics. The photons involved in these energy non-conserving processes are interpreted as virtual, and since for their existence they rely on the presence of the dipole, they are sometimes hypothesised as forming a localised cloud surrounding the ``bare" dipole \cite{cohen-tannoudji_photons_1989,compagno_atom-field_1995,compagno_dressed_1988,compagno_dressed_1990,compagno_detection_1988,compagno_bare_1991,passante_cloud_1985}. 

In both classical and quantum theories electromagnetic energy is supposed to come in two varieties; energy permanently tied to the charged source, and energy able to detach itself in the form of radiation. It would be natural to interpret the energy of the virtual cloud as the quantum version of the non-radiative classical energy, and to interpret real photons as those that contribute to the quantum version of the classical radiated energy. Since photons in the quantum theory are defined in terms of the operator-valued Maxwell fields, one can investigate the quantum versions of the classical radiation fields to determine whether or not these straightforward interpretations can in fact be made. To this end one must be able to identify the radiation fields in terms of photonic operators. This is achieved within the present paper. Our subsequent analysis of the quantum radiated energy flux of an excited dipole, which is defined in terms of the quantum Poynting vector, reveals that there is no justification for interpreting the virtual photon cloud as something that is localised around the dipole.

Only when the radiated energy flux is time-averaged does it coincide with the usual quantum optical energy flux of spontaneous emission. The time-average removes the contributions resulting from interactions with the virtual cloud, that occur over very short time scales. Thus, while the interpretation of the bare dipole as being surrounded by a virtual photon cloud from which photons are continually emitted and reabsorbed is still tenable, this virtual cloud must be understood as extending infinitely far from the source. Therefore, the conventional ``dressed" atom, which consists of the bare atom plus its virtual cloud is a highly non-local object.

We remark before continuing that there are other definitions of radiation in classical electrodynamics besides the definition given in terms of the ``radiation" fields, and the appropriate definition is still occasionally the subject of debate. Here we introduce quantum optical dipole ``radiation" fields, which vary as $1/x$ away from the source at the origin ${\bf 0}$. We then analyse the expectation value of the corresponding ``radiation" Poynting vector. The implications of our analysis for quantum electrodynamics are largely independent of whether or not one deems this Poynting vector to be a true representation of radiated energy flux.

\section{Quantum radiation fields}

Throughout the remainder of this paper we use natural units such that $\hbar=c=\epsilon_0 = \mu_0=1$ and $e^2=4\pi \alpha$ with $\alpha$ the fine structure constant. We consider the case of a non-relativistic stationary dipole consisting of a single charge $-e$ with mass $m$ anchored to the origin via an external potential describing a charge $+e$ fixed at the origin. The dipole is described by a non-relativistic quantum Schr\"odinger matter field $\psi$ satisfying the anti-commutation relation $\{\psi({\bf x}),\psi^\dagger({\bf x}')\}=\delta({\bf x}-{\bf x}')$ and with all other anti-commutators zero. The dipole interacts with the electromagnetic field, which is described by quantum transverse canonical Maxwell fields ${\bf A}_{\rm T}$ and $-{\bf E}_{\rm T}$ satisfying the commutation relation $[{\bf A}_{\rm T}({\bf x}), {\bf E}_{\rm T}({\bf x}')]=-i\delta_{ij}^{\rm T}({\bf x}-{\bf x}')$ where $\delta_{ij}^{\rm T}$ denotes the transverse delta-function. All other commutators between the Maxwell fields are zero. 

The minimal-coupling, i.e., Coulomb gauge Hamiltonian describing the dipole-field system reads
\begin{align}\label{h}
H= \int d^3x \, \psi^\dagger ({\bf x}){1\over 2m}\left[-i\nabla+e{\bf A}_{\rm T}({\bf 0})\right]^2\psi({\bf x}) +V + {1\over 2}\int d^3 x \, \left[{\bf E}_{\rm T}({\bf x})^2 + {\bf B}({\bf x})^2\right]
\end{align} 
where $V=V_{\rm ext}+V_{\rm self}$ includes both the external binding potential centered at the origin, and the dipole self-energy, which are respectively defined by
\begin{align}
V_{\rm ext} = \int d^3x \, \psi^\dagger({\bf x})\left[{-e^2\over 4\pi x}\right]\psi({\bf x}),\qquad V_{\rm self} = {1\over 2}\int d^3x \,{\bf E}_{\rm L}({\bf x})^2
\end{align}
with $\nabla \cdot {\bf E}_{\rm L}({\bf x}) = -e\psi^\dagger({\bf x})\psi({\bf x})$. The Magnetic field in (\ref{h}) is defined as usual by ${\bf B}({\bf x}) = \nabla\times {\bf A}_{\rm T}({\bf x})$.

The quantum Heisenberg equations of motion obtained from (\ref{h}) are the dipole approximated quantum Maxwell equations and quantum Lorentz-force law for the dipole acceleration operator. When solved these equations yield the retarded dipole electric source field
\begin{align}\label{eso}
{\bf E}^S(t,{\bf x}) ={\bf E}^S_{\rm T}(t,{\bf x})+{\bf E}_{\rm L}(t,{\bf x})
={1\over 4\pi x^3}\left\{3{\hat {\bf x}}\left[{\hat {\bf x}}\cdot {\bf d}(t_r)\right]- {\bf d}(t_r)\right\}+{1\over 4\pi x^2}\left\{3{\hat {\bf x}}\left[{\hat {\bf x}}\cdot {\dot {\bf d}}(t_r)\right] - {\dot {\bf d}}(t_r)\right\} +{1\over 4\pi x} {\hat {\bf x}}\times \left[{\hat {\bf x}}\times {\ddot {\bf d}}(t_r)\right]
\end{align}
with $t_r=t-x$ denoting the retarded time (with $c=1$) and with
\begin{align}
{\bf d}(t) = \int d^3x \, \psi^\dagger(t,{\bf x})[-e{\bf x}]\psi(t,{\bf x}).
\end{align}

The corresponding radiation source field is defined as the component of the source field in (\ref{eso}) that varies as $1/x$, which coincides with the component that depends on the dipole acceleration;
\begin{align}\label{es}
{\bf E}_{\rm rad}^S(t,{\bf x}) := {1\over 4\pi x} {\hat {\bf x}}\times \left[{\hat {\bf x}}\times {\ddot {\bf d}}(t_r)\right].
\end{align}
The magnetic radiation source field is then given by
\begin{align}\label{bs}
{\bf B}^S_{\rm rad}(t,{\bf x}) := {\hat {\bf x}}\times {\bf E}_{\rm rad}^S(t,{\bf x}) = -{1\over 4\pi x}{\hat {\bf x}}\times {\ddot {\bf d}}(t_r).
\end{align}

Since according to (\ref{es}) and (\ref{bs}) the radiation source fields are orthogonal to ${\bf x}$, we make an  ansatz for the total quantum radiation fields defined in terms of photonic operators by identifying the components of the usual mode expansions of the quantum Maxwell fields that are orthogonal to ${\bf x}$ in the following sense
\begin{subequations}\label{modeex}
\begin{align}
{\bf A}_{\rm T,rad}(t,{\bf x}):=&{i\over 2x}\sum_{\lambda=1,2} {\bf e}_\lambda({\bf x})\int_0^\infty d\omega\, \sqrt{\omega\over \pi}\left[a^\dagger_\lambda(t,\omega{\hat {\bf x}}) e^{-i\omega x}-a_\lambda(t,\omega{\hat {\bf x}})e^{i\omega x}\right], \\ 
{\bf E}_{\rm rad}(t,{\bf x}):=&{1\over 2x}\sum_{\lambda=1,2} {\bf e}_\lambda({\bf x})\int_0^\infty d\omega\, \sqrt{\omega^3\over \pi} \left[a_\lambda(t,\omega{\hat {\bf x}})e^{i\omega x} + a^\dagger_\lambda(t,\omega{\hat {\bf x}}) e^{-i\omega x}\right],\\
{\bf B}_{\rm rad}(t,{\bf x}):=&{1\over 2x}\sum_{\lambda=1,2} {\hat {\bf x}}\times{\bf e}_\lambda({\bf x})\int_0^\infty d\omega\, \sqrt{\omega^3\over \pi} \left[a_\lambda(t,\omega{\hat {\bf x}})e^{i\omega x} + a^\dagger_\lambda(t,\omega{\hat {\bf x}}) e^{-i\omega x}\right]
\end{align}
\end{subequations}
where $\{{\bf e}_1({\bf x}),{\bf e}_2({\bf x}),{\hat {\bf x}}\}$ composes a right-handed orthonormal triad of 3-vectors. The operators $a_\lambda(\omega{\hat {\bf x}})$ and $a^\dagger_\lambda(\omega{\hat {\bf x}})$ in (\ref{modeex}) are respectively annihilation and creation operators for a photon with polarisation $\lambda$ and wavevector ${\bf k}=\omega{\hat {\bf x}}$. They act within the continuous Bosonic Fock space built out of the single-particle space that consists of square-integrable ${\mathbb C}^2$-valued functions. Their relevant algebraic properties are specified entirely through the commutation relation
\begin{align}
[a_\lambda({\bf k}) ,a^\dagger_{\lambda'}({\bf k}')]=\delta_{\lambda\lambda'}\delta({\bf k}-{\bf k}').
\end{align}
The definitions in (\ref{modeex}) evaluate the photonic operators within a restricted space of wavevectors ${\bf k}=\omega{\hat {\bf x}}$ having variable magnitude $\omega \in (0,\infty)$, but which necessarily point in the direction from the source (at ${\bf 0}$) to the field point ${\bf x}$.

In order to justify the definitions in (\ref{modeex}) let us calculate the source components of these fields. The Heisenberg equation for the photon annihilation operator with general wavevector ${\bf k}$ is found using (\ref{h}), and can be integrated to yield the solution
\begin{align}\label{acoul}
a_\lambda(t,{\bf k}) = a_\lambda(0,{\bf k})e^{-i\omega t} + ig \int_0^t dt' e^{-i\omega (t-t')}{\dot {\bf d}}(t')\cdot{\bf e}_\lambda({\bf k})
\end{align}
where $\omega = |{\bf k}|$ and $g={1/\sqrt{2\omega (2\pi)^3}}$. Note that this equation holds for any ${\bf k}$ and therefore in particular when ${\bf k}=\omega{\hat {\bf x}}$. The integral term represents the source component, whose substitution into (\ref{modeex}) with ${\bf k}=\omega{\hat {\bf x}}$ yields the radiation source potential
\begin{align}\label{vecrad}
{\bf A}^S_{\rm T,rad}(t,{\bf x}) =&  {1\over (2\pi)^2 x} \int^t_0 dt' \sum_{\lambda =1,2} {\bf e}_\lambda({\bf x}) [{\bf e}_\lambda({\bf x})\cdot {\dot {\bf d}}(t')]{1\over 2}\int_0^\infty d\omega \,  \left[e^{i\omega(t'-t_r)}+e^{-i\omega(t'-t_r)}\right] \nonumber \\ =& {1\over (2\pi)^2 x} \int^t_0 dt' \sum_{\lambda =1,2} {\bf e}_\lambda({\bf x}) [{\bf e}_\lambda({\bf x})\cdot {\dot {\bf d}}(t')] \pi \delta(t'-t_r) \nonumber \\
=& -{1\over 4\pi x} {\hat {\bf x}}\times \left[{\hat {\bf x}}\times {\dot {\bf d}}(t_r)\right].
\end{align}

The radiation source fields can be obtained from (\ref{vecrad}) via
\begin{align}
{\bf E}^S_{\rm rad}(t,{\bf x}) = -{\dot {\bf A}}^S_{\rm T,rad}(t,{\bf x}),\qquad {\bf B}^S_{\rm rad}(t,{\bf x}) = -{\hat {\bf x}} \times {\dot {\bf A}}^S_{\rm T,rad}(t,{\bf x}),
\end{align}
or directly from (\ref{modeex}) as
\begin{subequations}\label{rads}
\begin{align}
{\bf E}^S_{\rm rad}(t,{\bf x}) =& {1\over (2\pi)^2x}\int_0^t dt' \sum_{\lambda=1,2} {\bf e}_\lambda({\bf x})\left[{\bf e}_\lambda({\bf x}) \cdot {\dot {\bf d}}(t')\right]{i\over 2}\int_0^\infty d\omega\, \omega\left[e^{i\omega(t'-t_r)}-e^{-i\omega(t'-t_r)}\right] \nonumber \\ =& {1\over (2\pi)^2 x} \int^t_0 dt' \sum_{\lambda =1,2} {\bf e}_\lambda({\bf x}) [{\bf e}_\lambda({\bf x})\cdot {\dot {\bf d}}(t')] \pi {d\over dt'} \delta(t'-t_r) \nonumber \\
=& {1\over 4\pi x} {\hat {\bf x}}\times \left[{\hat {\bf x}}\times {\ddot {\bf d}}(t_r)\right],\label{rads1} \\  {\bf B}^S_{\rm rad}(t,{\bf x})=&{\hat {\bf x}}\times {\bf E}^S_{\rm rad}(t,{\bf x}).
\end{align}
\end{subequations}

The radiation fields can also be found using the Poincar\' e gauge (multipolar) Hamiltonian leading to the same result as in (\ref{rads}). Thus, unlike the total fields (c.f. \cite{power_time_1999-1,power_time_1999}), the source components of the radiation fields are identical in either gauge. The source components of the total fields differ between the two gauges only by a static nearzone polarisation field which plays no role in radiative effects within the radiation zone. In fact, within the dipole approximation this polarisation field is entirely localised at the origin.

%One interesting property of the radiation fields is that unlike the corresponding total fields they satisfy the canonical commutation relation
%\begin{align}
%[E_{\rm rad,\lambda}({\bf x}),A_{\rm T,rad,\lambda'}({\bf x}')] = i\int_0^\infty d\omega d\omega'
%\end{align}

%allow one to define the configuration space radiation ``annihilation operator"
%\begin{align}
%b_\lambda({\bf x}) = {1\over \sqrt{2x}} \left[E_{\rm rad,\lambda}({\bf x}) +i x A_{\rm T,rad,\lambda}({\bf x})\right]
%\end{align}
%which satisfies
%\begin{align}
%.
%\end{align}
%One can therefore define single-particle position states via $\ket{\bf x} = b^\dagger({\bf x})\ket{0}$, which posses the desired normalisation property $\langle{\bf x}|{\bf x}'\rangle = \delta({\bf x}-{\bf x}')$. One can then tentatively interpret $|\langle {\bf x} | \psi \rangle|^2 d^3 x$ as the probability to detect a photon with state $\ket{\psi}$ within the volume element $d^3 x$ located at ${\bf x}$. There are some drawbacks to this interpretation, which does not (miraculously) solve the relativistic localisation problem. Firstly there is the use of the dipole approximation, and secondly the operator $b(t)$ does not commute with itself at spacelike separations, because it is defined in terms of the non-local field ${\bf A}_{\rm T}$.

%***LORENTZ TRANSFORMATION PROPERTIES?***
\section{The classification of photons}

The $S$-matrix gives the probability amplitudes associated with transitions occurring over infinite time intervals between the eigenstates $\{\ket{n}\}$ of $H_0$; $H_0\ket{n} =\omega_n \ket{n}$. The $S$-matrix element describing a transition from a bare state $\ket{i}$ with energy $\omega_i$ into a bare state $\ket{f}$ with energy $\omega_f$ has the form \cite{roman_advanced_1965,cohen-tannoudji_atom-photon_1992}
\begin{align}\label{Sfi}
S_{fi} =\delta_{fi} - 2\pi i \lim_{\tau\to \infty} \delta^\tau(\omega_f-\omega_i)T_{fi}
\end{align}
where $T$ is called the transition matrix, and is defined in terms of the Hamiltonian resolvent. The exact form of a transition matrix element is unimportant here. The important element of (\ref{Sfi}) for our purposes is the function
\begin{align}
\delta^\tau(\omega_f-\omega_i)= {1\over \pi}{\sin[(\omega_f-\omega_i)\tau/2]\over \omega_f-\omega_i},\qquad \lim_{\tau\to \infty} \delta^\tau(\omega_f-\omega_i) = \delta(\omega_f-\omega_i).
\end{align}
The presence of the delta function in (\ref{Sfi}) shows that $S$-matrix elements describe processes conserving $H_0$. Photons in QED are the quanta associated with the free field energy $H_F$. The distinction between real and virtual photons depends on the process under study. Photons occurring in processes that conserve $H_0$ are necessarily real, while others may be virtual. Of course, energy conserving processes can be made up of shorter energy non-conserving processes.

The above classification of photons is somewhat limited, because a process, in the sense used above, refers to a transition occurring over an infinite period of time. For interactions occurring over finite times the distinction between real and virtual photons is less straightforward and generally relies upon the use of the energy-time uncertainty relation. This uncertainty relation can be derived by noting that when the limit $\tau\to \infty$ is avoided in (\ref{Sfi}) one obtains only an approximate delta function $\delta^\tau(\omega_f-\omega_i)$ with a dominant peak at $\omega_f=\omega_i $ that has a width of the order of $1/\tau$. This is taken as expressing the conservation of $H_0$ through the $\ket{i}\to \ket{f}$ transition, to within an uncertainty of the order of $1/\tau$. We therefore obtain the energy-time uncertainty relation \cite{cohen-tannoudji_atom-photon_1992}
\begin{align}\label{et}
\Delta\omega \sim {1\over \tau}.
\end{align}

It is important to recognise that since time is not represented by an operator in conventional quantum theory, the energy-time uncertainty relation is not the same as the Heisenberg uncertainty principle, which necessarily holds for Fourier-conjugate operators like the position and momentum operators encountered in non-relativistic wave-mechanics. The quantitative justification for the energy-time uncertainty relation (\ref{et}) is based entirely on the presence of the function $\delta^\tau$ in (\ref{Sfi}). But when the limit $\tau\to \infty$ in (\ref{Sfi}) is avoided, as in the derivation of (\ref{et}), the justification for assuming that a transition probability amplitude has the structure given by the right-hand-side of (\ref{Sfi}) is less clear. Furthermore, the function $\delta^\tau$ exhibts oscillations outside of the main peak at $\omega_f=\omega_i$. As such there is nothing to prohibit nonzero values of $\omega_f$ outside the range $\omega_i\pm{1/ \tau}$, from contributing to the probability amplitude of a transition into a range of final states. This has the result that probability amplitudes for finite-time transitions into a complete set of final states become frequency cut-off dependent. The contributions of frequencies $\omega_f$ that do not obey (\ref{et}) can be significant for quite natural choices of cut-offs, for particular forms of interaction Hamiltonian (coupling strengths) \cite{stokes_extending_2012}. The energy-time uncertainty relation should not therefore be regarded as a fundamental principle that cannot be violated, but rather as a tool with which useful interpretations might be offered.

Of course, whether or not a so-called virtual photon is in fact real depends on whether or not it is possible to detect such a photon in the lab. The idea of detecting the virtual cloud directly has received some attention in the past, and has been modeled via direct coupling to a pointer \cite{compagno_detection_1988,compagno_bare_1991}. Based on the energy-time uncertainty principle, it has been suggested that measurements with duration $\tau< 1/(\omega_0 + \omega)$ where $\omega$ denotes the frequency of the photon and $\omega_0$ denotes the frequency of the dipole transition, are capable of resolving virtual emission events. In such measurements the dipole would therefore be perceived as bare by the pointer \cite{compagno_detection_1988,compagno_bare_1991}. The measurement process is capable of transferring an energy no larger than $1/\tau$ into the dipole-field system, so that if $\tau< 1/(\omega_0 + \omega)$ the virtual excitation becomes real. According to these ideas the distinction between real and virtual photons can not be made independent of the experimental setup considered. Moreover, as one might expect, no experiment could possibly distinguish between a real and virtual photon, because a detected photon is necessarily real.

Nevertheless, it is clear that the classification of photons into real and virtual groups always involves the free energy $H_0$, along with the question of its conservation. In contrast, in classical electrodynamics the quantity $H_0$ that results from a canonical formulation is not usually viewed as possessing any special significance. Instead the Maxwell fields are analysed directly. As pointed out in the introduction, the quantum Maxwell equations can formally be solved in the same way as in classical theory, and it is well known that virtual contributions to the fields cannot be neglected without sacrificing their causal nature \cite{biswas_virtual_1990,hegerfeldt_causality_1994,buchholz_there_1994,milonni_photodetection_1995,power_analysis_1997,sabin_fermi_2011,stokes_noncovariant_2012}. This shows that while not directly observable virtual photons do have physical significance beyond their role played in level-shift phenomena.

The retarded fields can be split into near zone, intermediate zone and far zone (radiation) fields each of which is fully retarded. This suggests that the radiation fields in particular contain virtual contributions, in which case the virtual cloud could not possibly be considered as something localised around the atom. On the other hand the localisation of the virtual cloud is qualitatively consistent with the energy-time uncertainty relation if we assume that virtual photons propagate with the finite speed of light. This would mean that they cannot travel as far from the dipole as real photons, because they exist on much shorter timescales. Such reasoning is however, dubious, given that photons cannot generally be localised, and the sense in which they can be viewed as propagating is unclear.

In the following section we {\em define} the real contribution to the radiated energy-flux as that which gives rise to the standard quantum optical spontaneous emission rate usually calculated using first order $S$-matrix theory. This is the rate at which the excited dipole makes a transition into a lower energy state and in doing so emits a photon with exactly the frequency of the transition. The emitted photons are therefore necessarily real. We define the remaining contribution to the energy-flux as the virtual contribution for which emission events do not necessarily conserve the free energy, despite by assumption satisfying an energy-time uncertainty relation of the kind in (\ref{et}). We can then ask whether or not the radiation fields introduced in the previous section contribute to the virtual component of the energy-flux, which if the case would mean that the virtual contribution does not vanish infinitely far from the charged source.

\section{Quantum radiated energy flux}

In \cite{power_quantum_1992,salam_molecular_2008,salam_molecular_2009} the spontaneous emission rate of an excited dipole is calculated by evaluating the expectation value of the associated radiated power using second order perturbation theory. The radiated power is defined in terms of the quantum Poynting vector. This method of deriving the spontaneous emission rate has direct physical appeal, because it determines the relationship between two fundamental quantities. The Poynting vector represents the electromagnetic energy-flux derived from first principles via Noethers theorem, while the rate of spontaneous emission calculated using first order $S$-matrix theory is a fundamental result in quantum electrodynamics.

The first step in the calculation is the determination of the expected Poynting vector, the result of which can be expressed in the form
\begin{align}\label{S}
\langle {\bf S}_{0;e} \rangle ={1\over 2}( \langle{\bf E}\times{\bf B}\rangle_{0;e} - \langle {\bf B} \times {\bf E}\rangle_{0;e} ) = \langle {\bf S}^{\rm real}\rangle_{0;e}  + \langle {\bf S}^{\rm virtual} \rangle_{0;e}
\end{align}
where the expectation value is taken in the photon vacuum $\ket{0}$ and with the dipole in an excited state $\ket{e}$. In (\ref{S}) the terms $\langle {\bf S}^{\rm real}\rangle$ and $\langle {\bf S}^{\rm virtual} \rangle$ give the real and virtual contributions respectively, whose explicit form follows from lengthy calculations. Their identification as real and virtual contributions is motivated using Poynting's theorem, which gives the energy-flux across the surface of a sphere with radius $x$ as
\begin{align}\label{P}
P = \int d\Omega \,  x^2\, {\hat {\bf x}} \cdot \langle{\bf S} \rangle_{0;e}
\end{align}
where $d\Omega$ denotes integration over the unit sphere. Substituting the explicit expression found for $\langle{\bf S}^{\rm real}\rangle$ into (\ref{P}) gives the real component of the energy-flux as
\begin{align}\label{Pradem}
P^{\rm real} =  \int d\Omega \,  x^2\, {\hat {\bf x}} \cdot \langle{\bf S}^{\rm real} \rangle_{0;e}= \sum_{m<e} {\omega_{em}^4 |{\bf d}_{em}|^2\over 3\pi} = \sum_{m<e} \Gamma_{e\to m}\omega_{em},~~~~~~\Gamma_{e \to m} := {\omega_{em}^3 |{\bf d}_{em}|^2 \over 3\pi},
\end{align}
in which we recognise $\Gamma_{e\to m}$ as the standard quantum optical spontaneous emission rate for the transition $\ket{e}\to \ket{m}$. The interpretation of $P^{\rm real}$ is clear; (real) photons with the energy $\omega_{em}$ of the dipole transition $\ket{e}\to \ket{m}$ are emitted at a rate $\Gamma_{e\to m}$, giving an energy flux $\omega_{em}\Gamma_{e\to m}$. The total real energy flux is the sum over all transitions for which $e>m$, i.e., for which the emission events conserve the canonical free energy. This is the justification for identifying ${\bf S}^{\rm real}$ as the component of the Poynting vector that gives the contribution of real photons.

So, when virtual contributions are ignored the expected Poynting vector gives the standard spontaneous emission flux of an excited dipole usually found using Fermi's golden rule, i.e., first order $S$-matrix theory. Despite the simplicity of the final result (\ref{Pradem}) the calculations in \cite{power_quantum_1992,salam_molecular_2008,salam_molecular_2009} are fairly involved owing to the use of the complete Maxwell fields, which result in complicated mode function summations.

Here we repeat the calculation carried out in \cite{power_quantum_1992,salam_molecular_2008,salam_molecular_2009}, but instead of using the total fields we use the radiation fields introduced in (\ref{modeex}). We employ the same methodology of using time-dependent perturbation theory in the Heisenberg picture to find the expectation value
\begin{align}\label{flux}
\langle {\bf E}_{\rm rad}\times {\bf B}_{\rm rad}\rangle_{0;e}= {\hat {\bf x}}\langle{{\bf E}_{\rm rad}}^2 \rangle_{0;e}.
\end{align}
We then substitute the result into
\begin{align}\label{Prad}
P_{\rm rad} = \int d\Omega \,  x^2\, {\hat {\bf x}} \cdot\langle {\bf S}_{\rm rad} \rangle_{0;e} =\int d\Omega \,  x^2\, \langle {\bf E}^2_{\rm rad} \rangle_{0;e}
\end{align}
to obtain the radiated energy flux. We find that when virtual contributions are ignored we obtain precisely the same result as is obtained using the total fields; $P_{\rm rad}=P^{\rm real}$ with $P^{\rm real}$ given in (\ref{Pradem}). This shows that only the radiation fields contribute to the real spontaneous emission of photons. Moreover, for the purpose of calculating the spontaneous emission rate given by (\ref{Pradem}) our method offers a significant simplification over those in \cite{power_quantum_1992,salam_molecular_2008,salam_molecular_2009}, because the radiation field mode expansions in (\ref{modeex}) turn out to be much simpler to work with than the corresponding mode expansions of the total fields. Since the near zone and intermediate zone components do not contribute to the final result (\ref{Pradem}), there is no need to retain them throughout the calculation. Of course, for any phenomena such that the near and intermediate zone fields do make a contribution, use of the radiation fields in (\ref{modeex}) could at best be viewed as an approximation. Our calculation demonstrates that spontaneous emission is not an example of such a phenomenon.

Our calculation further shows that the radiation fields \emph{do} contribute to the remaining virtual component of the energy-flux $P^{\rm virtual}$. That is, we do not obtain $P_{\rm rad} \equiv P^{\rm real},~P_{\rm rad}^{\rm virtual} \equiv 0$, meaning that just like in the calculation involving the total fields one still has to neglect virtual contributions in order to obtain the standard real spontaneous emission rate from the Poynting vector \cite{power_quantum_1992,salam_molecular_2008,salam_molecular_2009}. Thus, the virtual contributions to the energy-flux are not produced by the near and intermediate zone fields alone, and a virtual component of the radiated energy flux \emph{does} survive infinitely far from the dipole source. Starting from the radiation fields in (\ref{modeex}), no additional approximations are used in this section, apart from the dipole approximation and perturbation theory.

We use the Poincar\'e gauge (multipolar) Hamiltonian in the electric dipole approximation
\begin{align}\label{h2}
H=H_0-{\bf d}\cdot {\bf D}_{\rm T}({\bf 0})
\end{align}
where ${\bf D}_{\rm T}$ denotes the transverse displacement field and
\begin{align}
H_0 = \sum_n \omega_n b^\dagger_n b_n +\int d^3k \sum_\lambda \omega\left[a^\dagger_\lambda({\bf k}) a_\lambda({\bf k})+{1\over 2}\right].
\end{align}
We have neglected any self-energy terms of order $e^2$ in (\ref{h2}), because these terms do not contribute in the second order evaluation of the energy flux in (\ref{flux}). The operator $b^\dagger_n$ in (\ref{h2}) creates the dipole energy eigenstate $\ket{n}$ from the vacuum; $b^\dagger_n\ket{0}=\ket{n}$, and is related to the field $\psi$ via
\begin{align}
b^\dagger_n := \int d^3 x\, \varphi_n({\bf x}) \psi^\dagger({\bf x})
\end{align}
where $\varphi_n$ is a single-particle bare dipole energy eigenfunction defined through the equation
\begin{align}
\left[{-\nabla^2\over 2m}-{e^2\over 4\pi x}\right]\varphi_n({\bf x}) = \omega_n\varphi_n({\bf x}).
\end{align}
These energy basis operators can be used just like the position basis fields $\psi$ and $\psi^\dagger$ to express dipole operators. For example, the dipole moment can be written
\begin{align}\label{d2}
{\bf d}(t) = \sum_{nm} {\bf d}_{nm}b^\dagger_n(t)b_m(t)
\end{align}
where ${\bf d}_{nm} := \bra{n}{\bf d}(0)\ket{m}$.

For our perturbative calculation we introduce the interaction picture operators $\alpha_\lambda(t,{\bf k})$ and $\beta_n(t)$ defined by
\begin{align}\label{intpic}
\beta_n(t):=e^{i\omega_n t}b_n(t),\qquad \alpha_\lambda(t,{\bf k}):=e^{i\omega t}a_\lambda({\bf k}).
\end{align}
The integrated equations of motion for these operators are \cite{salam_molecular_2008}
\begin{subequations}
\begin{align}
\alpha_\lambda(t,{\bf k}) =& \alpha_\lambda(0,{\bf k}) + \omega g \sum_{n,m}{\bf e}_\lambda({\bf k}) \cdot {\bf d}_{nm}\int_0^t dt'\, e^{i(\omega_{nm}+\omega)t'}\beta^\dagger_n(t')\beta_m(t'), \\
\beta_n(t)=& \beta_n(0)-\int d^3 k \,\omega g \sum_m \sum_\lambda  {\bf e}_\lambda({\bf k})\cdot {\bf d}_{nm}\int_0^t dt' \, e^{-i\omega_{nm}t'}\beta_m(t')\left[e^{-i\omega t'}\alpha_\lambda(t',{\bf k})-{\rm H.c}\right]
\end{align}
\end{subequations}
where $\omega_{nm} := \omega_n-\omega_m$. The zeroth order terms in the perturbative expansions of the above solutions are simply the free components
\begin{align}\label{zeroth}
\alpha^{(0)}_\lambda(t,{\bf k})=\alpha_\lambda(0,{\bf k})=:\alpha_\lambda({\bf k}),\qquad \beta_n^{(0)}(t) = \beta_n(0)=:\beta_n.
\end{align}

For a second order (in $e$) evaluation of the expectation value in (\ref{flux}) we require the radiation field ${\bf E}_{\rm rad}$ upto second order. The zeroth order field is simply the component giving the free evolution;
\begin{align}\label{0}
{\bf E}_{\rm rad}^{(0)}(t,{\bf x}):=&{1\over 2x}\sum_{\lambda=1,2} {\bf e}_\lambda({\bf x})\int_0^\infty d\omega\, \sqrt{\omega^3\over \pi} \left[\alpha_\lambda(\omega{\hat {\bf x}})e^{-i\omega t_r} + \alpha^\dagger_\lambda(\omega{\hat {\bf x}}) e^{i\omega t_r}\right].
\end{align}
The first order field is easily obtained by substituting the zeroth order approximation of $\beta_n(t)$ given in (\ref{zeroth}) and its conjugate into (\ref{d2}), and then substituting the result into (\ref{rads1}) to give
\begin{align}\label{1}
{\bf E}^{(1)}_{\rm rad}(t,{\bf x}) = {1\over 4\pi x} \sum_\lambda \sum_{nm} {\bf e}_\lambda({\bf x})  [{\bf e}_\lambda({\bf x})\cdot {\bf d}_{nm}]\omega_{nm}^2\beta_n^\dagger \beta_m e^{i\omega_{nm}t_r}.
\end{align}
Finally the second order field ${\bf E}_{\rm rad}^{(2)}(t,{\bf x})$ is given by
\begin{align}\label{2}
{\bf E}_{\rm rad}^{(2)}(t,{\bf x}):=&{1\over 2x}\sum_{\lambda=1,2} {\bf e}_\lambda({\bf x})\int_0^\infty d\omega\, \sqrt{\omega^3\over \pi} \left[\alpha^{(2)}_\lambda(t,\omega{\hat {\bf x}})e^{-i\omega t_r} + \alpha^{(2)\dagger}_\lambda(t,\omega{\hat {\bf x}}) e^{i\omega t_r}\right]
\end{align}
with
\begin{subequations}\label{alphbet2}
\begin{align}
\alpha^{(2)}_\lambda(t,{\bf k}) &= \omega g \sum_{mn} {\bf e}_\lambda({\bf k})\cdot {\bf d}_{mn}\int_0^t dt' \, e^{i(\omega_{mn}+\omega)t'}\left[\beta_m^\dagger \beta^{(1)}_n(t') + {\beta_m^{(1)}}^\dagger(t')\beta_n\right],\\
\beta^{(1)}(t) &= \int d^3k \, \omega g \sum_\lambda \sum_m {\bf e}_\lambda({\bf k}) \cdot {\bf d}_{nm} \beta_m\left[\alpha_\lambda({\bf k})f^*(\omega_{mn}+\omega,t) - \alpha^\dagger_\lambda({\bf k})f^*(\omega_{mn}-\omega,t) \right]
\end{align}
\end{subequations}
wherein for convenience we have defined the function
\begin{align}
f(\omega,t) := {e^{i\omega t}-1\over i\omega}.
\end{align}

Substituting the second order expansion of ${\bf E}_{\rm rad}$ into (\ref{flux}) we obtain the expected energy flux correct to second order, as a sum of three terms;
\begin{align}\label{pflux}
\langle {\bf E}_{\rm rad}^2\rangle_{0;e} =\langle {\bf E}^{(1)}_{\rm rad}\cdot {\bf E}^{(1)}_{\rm rad}\rangle_{0;e}+\langle {\bf E}^{(0)}_{\rm rad}\cdot {\bf E}^{(2)}_{\rm rad}\rangle_{0;e} +\langle {\bf E}^{(2)}_{\rm rad}\cdot {\bf E}^{(0)}_{\rm rad}\rangle_{0;e}.
\end{align}
The first term above gives the contribution from the first order fields and is easily evaluated using (\ref{1}) as
\begin{align}\label{1st}
\langle {\bf E}^{(1)}_{\rm rad}\cdot {\bf E}^{(1)}_{\rm rad}\rangle_{0;e} = {1\over 16\pi^2 x^2} \sum_\lambda \sum_m |{\bf e}_\lambda({\bf x}) \cdot {\bf d}_{em}|^2\omega_{em}^4,
\end{align}
which we recognise as an emission flux to which all dipole levels $n>e$ and $n<e$ contribute \cite{power_quantum_1992,salam_molecular_2008,salam_molecular_2009}.

In a similar fashion the remaining contributions in (\ref{pflux}) are found using (\ref{0}), (\ref{2}) and (\ref{alphbet2}) to be
\begin{align}
\langle {\bf E}_{\rm rad}^{(0)}&({\bf x})\cdot {\bf E}_{\rm rad}^{(2)}({\bf x})\rangle_{0;e} \nonumber \\ &= {1\over 4x^2(2\pi)^3}\sum_\lambda \sum_m|{\bf e}_\lambda({\bf x})\cdot {\bf d}_{em}|^2\int_0^\infty d\omega\, \omega^2e^{-i\omega t_r}\left[{d^2\over dt'^2}\left\{f^*(\omega_{em}-\omega,t')e^{i\omega_{em}t'}-f(\omega_{em}+\omega,t')e^{-i\omega_{em}t'}\right\}\right]_{t'=t_r}
\end{align}
with the remaining term the complex conjugate of the above. This expression can be partitioned into time-dependent and time-independent contributions. We interpret the time-dependent contributions as those arising from interactions with the virtual cloud, which will be justified in what follows. Meanwhile, the time-independent contribution is
\begin{align}\label{res}
\langle {\bf E}_{\rm rad}^{(0)}\cdot &{\bf E}_{\rm rad}^{(2)} \rangle_{0;e}^{\rm real} = {i\over 4x^2 (2\pi)^3}\sum_\lambda \sum_m |{\bf e}_\lambda({\bf x})\cdot {\bf d}_{em}|^2\int d\omega \, \omega^4 \left[{1\over \omega - \omega_{em}}-{1\over \omega + \omega_{em}}\right].
\end{align}
To evaluate the integral above we use a retarded propagator prescription consistent with the use of the retarded fields. This entails adding an infinitesimal imaginary term $i\epsilon,~\epsilon>0$ to the denominators in square brackets in (\ref{res}), which shifts the simple poles into the lower half complex plane. Using the distributional identity
\begin{align}
\lim_{\epsilon\to 0}{1\over \omega +i\epsilon} = {{\mathcal P}\over \omega}-i\pi\delta(\omega)
\end{align}
with ${\mathcal P}$ denoting the Cauchy principal integral value, we then obtain
\begin{align}\label{res2}
\langle {\bf E}_{\rm rad}^{(0)}\cdot &{\bf E}_{\rm rad}^{(2)} \rangle_{0;e}^{\rm real} + {\rm c.c}={1\over 16\pi^2 x^2} \sum_\lambda \sum_m |{\bf e}_\lambda({\bf x}) \cdot {\bf d}_{em}|^2\omega_{em}^4{\rm sgn}(\omega_{em}).
\end{align}
Thus, summing up all contributions we obtain to second order in $e$
\begin{align}\label{Poyntf}
{\hat {\bf x}}\langle {\bf E}^2_{\rm rad} \rangle_{0;e} ={{\hat {\bf x}}\over 8\pi^2 x^2} \sum_\lambda \sum_{m<e} |{\bf e}_\lambda({\bf x}) \cdot {\bf d}_{em}|^2\omega_{em}^4 +{\hat {\bf x}}\langle {\bf E}_{\rm rad}^2 \rangle_{0;e}^{\rm virtual}.
\end{align}
To understand the significance of the first term above note that if we now substitute the first term in (\ref{Poyntf}) into (\ref{Prad}) and perform the integration over solid angle we obtain
\begin{align}\label{Pradem2}
P^{\rm real}_{\rm rad} = \sum_{m<e} {\omega_{em}^4 |{\bf d}_{em}|^2\over 3\pi} = \sum_{m<e} \Gamma_{e\to m}\omega_{em},~~~~~~\Gamma_{e \to m} := {\omega_{em}^3 |{\bf d}_{em}|^2 \over 3\pi},
\end{align}
which is identical to (\ref{Pradem}).

Having identified the contribution of real photons to the radiated power, the remaining contribution $P_{\rm rad}^{\rm virtual}$ arising from the second term ${\hat {\bf x}}\langle {\bf E}_{\rm rad}^2 \rangle_{0;e}^{\rm virtual}$ in (\ref{Poyntf}), must be the contribution of the virtual cloud, and is given by
\begin{align}\label{Pvirt}
P^{\rm virtual}_{\rm rad}(t) = {-i\over 4(2\pi)^3}\int d\Omega \sum_\lambda \sum_m \omega_{em}^2|{\bf e}_\lambda({\bf x}) \cdot {\bf d}_{em}|^2 \int_0^\infty d\omega \,\omega^2\left[\frac{e^{-i (\omega-\omega_{em}) t_r}}{\omega-\omega_{em}}-\frac{e^{-i (\omega+\omega_{em})t_r }}{\omega+\omega_{em}}\right]+{\rm c.c.}.
\end{align}
The virtual radiated power is clearly non-zero in general, but its time-dependence is entirely oscillatory. As a result the time-average of the time-dependent integrand in (\ref{Pvirt}) vanishes, so that
\begin{align}
{\overline P}^{\rm virtual}_{\rm rad} = \lim_{T\to \infty} {1\over T}\int_0^T dt\, P_{\rm rad}^{\rm virtual}(t) = 0.
\end{align} 
Since $P^{\rm real}$ is time-independent it follows that ${\overline P}_{\rm rad} = {\overline P}^{\rm real} = P^{\rm real}$. These time-averaged results are consistent with the interpretation of the virtual emission and reabsorbtion events as transient phenomena occurring over very short time-scales. The virtual radiative interactions appear to be characterised entirely through their temporal properties, and as such, their seems to be no good reason for interpreting the virtual cloud as something spatially localised around the dipole. Indeed the non-zero contribution $P^{\rm virtual}_{\rm rad}$ shows that virtual photon effects do extend infinitely far from the dipole source, despite their temporally transient behavior.

\section{Conclusions}

In this paper we have exhibited expressions for quantum dipole radiation fields in terms of photonic operators. The aim has been to assess whether real photonic energy can be identified as the quantum version of radiated electromagnetic classical energy, and whether virtual photonic energy can be identified as the quantum version of non-radiative classical electromagnetic energy. We have demonstrated that the answer to this question is negative. The underlying reason for this is a shift, when the transition to the quantum theory is made, in the means by which electromagnetic phenomena are interpreted. In quantum electrodynamics the concept of a \emph{process} in which photons are emitted, absorbed, and exchanged, plays a central role. This, no doubt, is the result of the general quantum field-theoretic strategy of perturbatively calculating scattering-matrix elements and cross-sections as the basic theoretical predictions.

%The nature of a process in quantum field theory is determined by whether or not it conserves the canonical free energy $H_0$, and accordingly, photons in QED are classified as real or virtual based on the extent to which the associated emission process conserves $H_0$. In classical electrodynamics $H_0$ plays no such privileged role, because the Maxwell equations can be solved directly and these solutions are then used to obtain physical predictions.

We have shown that the quantum dressed atom must be a highly non-local object, that extends infinitely far from the bare atom. We have shown that even when quantum radiation fields alone are used for the calculation of the Poynting vector energy-flux, we do not simply obtain the standard quantum optical spontaneous emission rate, which accounts for all emission events of photons that are necessarily real. This means that according to classical interpretive strategies of identifying radiation as the component of the energy-flux obtained from the radiation fields, both real and virtual photons must be involved in the emission of radiation, which is contrary to the typical quantum viewpoint whereby only real photons play a role.

Since one expects the classification of electromagnetic phenomena into radiative and non-radiative groups to be unique, one might expect that radiative classifications made in terms of free-energy should be the same as those made in terms of the spatial fields. As we have shown this is not the case. Moreover, it is important to note that either of these classification schemes can in principle be employed in either classical or quantum theoretical settings. Certainly, $H_0$ along with normal variables corresponding to photonic operators can be identified in the classical theory. Likewise, the spatial dependence of the source fields can be analysed in the quantum theory. Thus, in understanding radiative phenomena we have at least two inequivalent classification schemes (fields and photons) and at least two theoretical settings (classical and quantum), but the situation is not so simple so as to require that a given classification scheme must be paired with a given theoretical setting.

\bibliography{rads.bib}

\end{document}